# Nitrogen-Vacancy Nuclear Magnetic Resonance Spectroscopy of $^{15}N$ Spins in Potassium Nitrite Solution

Sehrish Iqbal,[1] Jitender Kumar,[1] Suvechhya Lamichhane,[2] Rupak Timalsina,[1] Sy-Hwang Liou,[2] and Abdelghani Laraoui[1,2*]

[1]Department of Mechanical & Materials Engineering, University of Nebraska-Lincoln. Lincoln, NE 68588, USA

[2]Department of Physics and Astronomy and the Nebraska Center for Materials and Nanoscience, University of Nebraska-Lincoln, Lincoln, NE 68588, USA

[*]Corresponding Author: alaraoui2@unl.edu

## ABSTRACT

Nitrogen is essential in agricultural fertilizers, yet its nanoscale interfacial dynamics remain largely unexplored. Although quadrupolar broadening complicates $^{14}N$ nuclear magnetic resonance (NMR) detection, $^{15}N$ (I=1/2) provides well-defined nuclear-spin signatures but remains challenging to detect at the nanoscale due to its low natural abundance and small gyromagnetic ratio. Here, we use a microfluidic quantum sensing platform based on shallow nitrogen vacancy (NV) centers in diamond to detect $^{15}N$ NMR from $KNO_3$ solution at room temperature and in a low magnetic field regime. Within femtoliter-scale sensing volumes, we measure a magnetic-field variance of up to ~442 nT², ~43 times greater than the expected statistical polarization. The $^{15}N$ spin correlation decays within ~500 μs, mainly limited by the NV spin lattice relaxation time and not by the molecular diffusion of the $KNO_3$ molecules. This is consistent with the formation of a $KNO_3$ adsorption layer on the diamond surface, supported by surface characterization.

**KEYWORDS:** *Nitrogen-vacancy, potassium nitrate, NMR, spin correlation, nitrogen*

Nitrogen (N) is an essential macronutrient for plant growth and a critical component of agricultural productivity, as it is required for the synthesis of proteins, nucleic acids, chlorophyll, and other biomolecules.[1] Nitrogen fertilizers are therefore widely used to sustain crop growth and enhance agricultural yields.[2,3] However, nitrogen in agricultural systems undergoes complex transformations between different chemical forms, with nitrate ion ($NO_3^-$) being one of the most mobile forms of nitrogen in soils and water. Excess nitrate can be transported through soil and water systems, contributing to nutrient losses and environmental impacts. Understanding and monitoring nitrogen diffusion at relevant spatial and chemical scales is consequently important for improving nitrogen-use efficiency and developing sustainable agricultural practices specifically at the liquid-solid interface i.e. capillaries in plants.[2,3]

Potassium nitrate ($KNO_3$) provides a well-defined and water-soluble source of nitrate and is commonly used as a nitrogen and potassium fertilizer, making it well suited for investigating nitrate at molecular and nanoscale levels. $KNO_3$ contains two naturally occurring nitrogen isotopes, with $^{14}N$ being the predominant isotope and $^{15}N$ present at much lower natural abundance (0.37%). $^{14}N$ is a spin-1 nucleus and its quadrupolar interaction can provide substantial line broadening, which can limit spectral resolution and quantitative detection.[4] In contrast, $^{15}N$ is a spin-1/2 isotope and does not experience quadrupolar broadening, resulting in narrower and better-resolved NMR resonances.[5] These characteristics make $^{15}N$ particularly advantageous for nanoscale NMR detection.

Conventional characterization techniques, such as nuclear magnetic resonance (NMR) spectroscopy, can provide detailed structural and compositional information.[4,6,7] However, they often require relatively large sample volumes and specialized instrumentation, high magnetic field and sometime need cryogenic temperatures to enhance polarization which can limit their applicability to nanoscale samples under practical experimental conditions. Consequently, the development of sensitive, local and rapid sensing strategies is important for detecting nitrogen from nanoscale volumes at room temperature and at low magnetic fields are highly desirable. In this context, recently nitrogen-vacancy (NV) center-based NMR offers a promising approach for nanoscale detection of nuclear spins with high spatial sensitivity under ambient conditions.[8,9] This new sensing platform, based on optical detection of electron spin resonances of NV centers in diamond, has emerged for measuring weak magnetic fields at the nanometer scale.[10–13] NV offers a combination of good spatial resolution and enhanced magnetic sensitivity[14,15] in a wide range of temperatures (0.3 – 600 K) and magnetic fields (up to a few T), opening up new opportunities in chemical[16] and condensed physics matter research.[17] NV-NMR was used to detect $^{13}C$ and $^{14}N$ spins in diamond,[18,19] $^{1}H$ protons[20,21] and $^{19}F$ nuclear spins[20] of solutions. By using quantum lock-in-detection,[22,23] NV-NMR with a spectral resolution of sub-Hz was achieved.[24,25]

Here, we use a microfluidic diamond quantum-sensing platform to detect $^{15}N$ NMR signals from 0.2 M $KNO_3$–deionized (DI) water solutions, spanning 10% $^{15}N$-enriched and naturally abundant $KNO_3$ (0.37% $^{15}N$), at room temperature and at moderate magnetic fields (20–43 mT). By using XY8-N-based spin-correlation spectroscopy we observe a long (400 – 600 μs) correlation decay time reaching the NV spin-lattice relaxation time ($T_1$) limit of the shallow NVs. Analysis of the time-dependent $^{15}N$ spin correlation signal enabled us to estimate the self-diffusion of $^{15}N$ nuclei near the diamond interface, revealing diffusion orders of magnitude slower than in bulk solution. We attribute this behavior to the formation of a $KNO_3$-containing interfacial layer, up to ~ 10 nm thick, during the solution flow. This interpretation is supported by atomic force microscopy (AFM) and X-ray photoelectron spectroscopy (XPS) of the diamond surface. The calibrated magnetic-field signal for the enriched solution is ~43 times larger than the theoretically expected statistical polarization signal within the detection volume. These results demonstrate the potential of microfluidic diamond quantum sensing for detecting $^{15}N$ nuclear spins in aqueous nitrate solutions and establish a foundation for nanoscale nitrogen sensing in chemical and fertilizer systems.

The schematic of the microfluidic NV-NMR setup used to detect $^{15}N$ in $KNO_3$ is illustrated in **Figure 1a** and discussed in detail in the Supporting Information Section S1 and **Figure S1**. A shallow NV layer was created by $^{15}N$ ion implantation into a (100)-oriented, cut, and polished, electronic grade diamond substrate ($3 \times 2 \times 0.1$ mm$^3$). See Reference (26) for more details. The implantation is followed by two hours temperature annealing at 1123 K under ultrahigh-vacuum conditions (~$10^{-6}$ Torr) and subsequent cleaning in a boiling tri-acid mixture.[26,27] A $KNO_3$ powder (ACS reagent, ≥ 99.0% purity) was purchased from Sigma Aldrich, and diluted in DI water with a concentration of 0.2 M. For this work, 0.2 M aqueous solutions of naturally abundant (0.37%) $KNO_3$ and 10% $^{15}N$-enriched $KNO_3$ were used for NV-NMR detection of $^{15}N$ nuclear spins.

The NV-NMR setup consists of a thick (0.5 mm) glass coverslip patterned with microwave (MW) transmission lines for coherent manipulation of the NV spins (**Figure S1**, Supporting Information). As shown in **Figure 1b**, the diamond is mounted directly on top of the MW lines using a UV-curable adhesive chosen for its strong bonding strength and negligible additional fluorescence.[28] The implanted NV layer is positioned close to the upper surface of the diamond, facing the microfluidic channel (**Figure 1a**), allowing direct interaction with nuclear spins of $KNO_3$/DI water solution. This configuration minimizes the risk of contaminating the NV layer with the UV adhesive while allowing the NV centers to directly interact with the flowing $KNO_3$/DI water solution. A 100-µm-deep microfluidic channel was fabricated in polydimethylsiloxane (PDMS) and positioned directly above the diamond sensing region. The PDMS channel was bonded to the glass MW platform using oxygen-plasma treatment, producing a robust and leak-tight interface suitable for continuous-flow measurements.

To characterize the shallow NV ensembles used for this study,[26,27] we first performed optically detected magnetic resonance (ODMR) measurements at room temperature. For (100)-oriented NV-ensemble diamond, four ODMR peaks ($f_-$ for $|0\rangle \leftrightarrow |-1\rangle$, and $f_+$ for ($|0\rangle \leftrightarrow |+1\rangle$, $f_-^m$ and $f_+^m$ for the merged other NV orientations) are observed at an external magnetic field ($B_{\text{app}}$) aligned along the [111] NV quantization axis. Here, we mainly focus on the $f_+$ ($|0\rangle \leftrightarrow |+1\rangle$) NV spin transition. A representative ODMR spectrum at $B_{\text{app}}$ of 40.96 mT at MW power of 0.32 W is plotted in **Figure S2a** (Supporting Information).

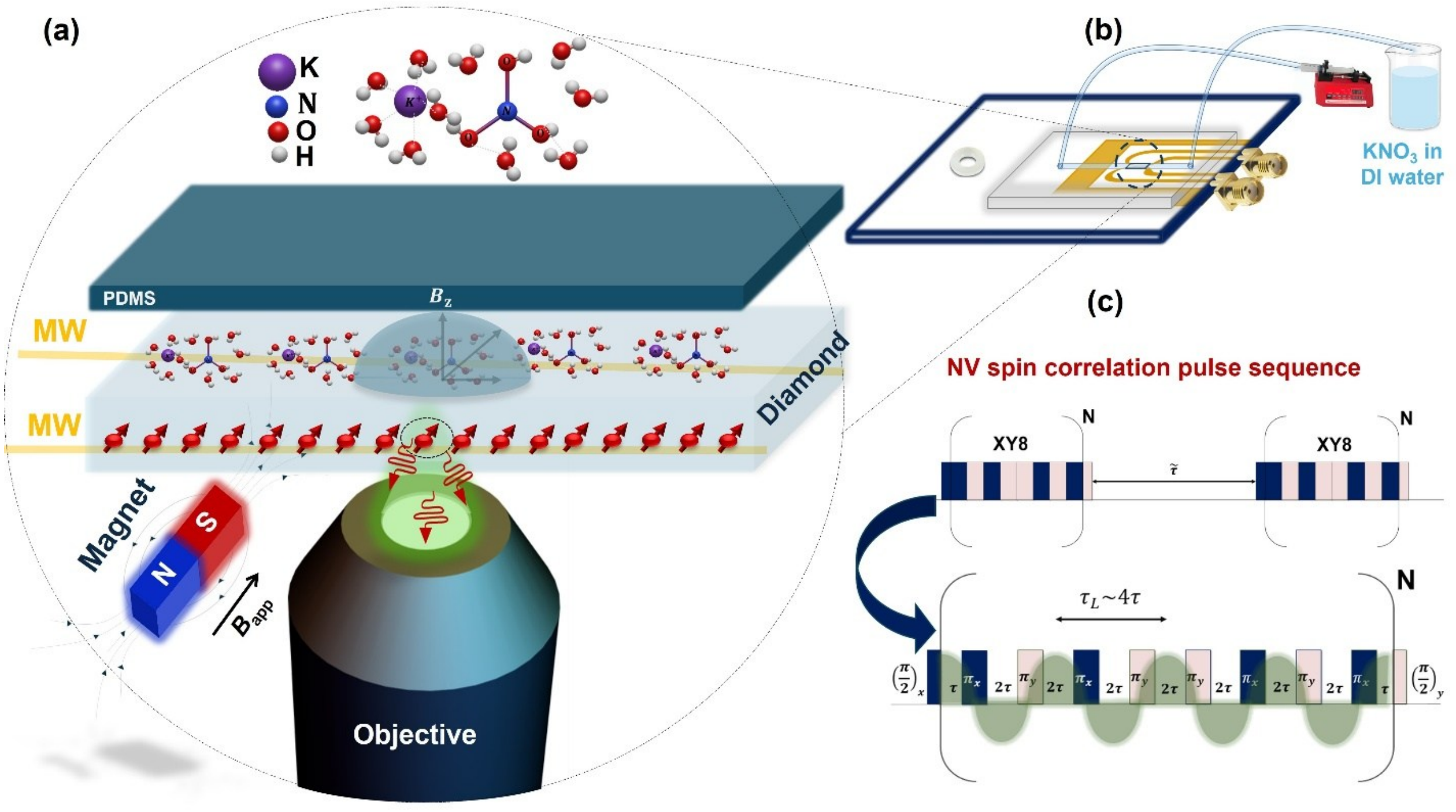

**Figure 1: (a)** Schematic of the NV-based microfluidic sensing platform. $KNO_3$ solution continuously flows through 100 µm-wide PDMS microchannels above the near-surface NV layer of the diamond while an external magnetic field is applied. **(b)** Device assembly showing the diamond sandwiched between the gold-plated Cu/Ta microwave PCB and the PDMS microfluidic channel with inlet and outlet tubing connected to a syringe pump. **(c)** NV-XY8-N spin correlation spectroscopy pulse sequence used for the NV-NMR study of $KNO_3$ solution.

We characterized the quantum coherence properties of the shallow NVs in bare diamond (no $KNO_3$ and water) and discussed it in Supporting Information Section S1. The measured NV $T_1$ is 1.499 ± 0.11 ms (**Figure S2c**, Supporting Information) and Hahn-echo spin-spin relaxation time, $T_2$, is 4.23 ±0.07 µs (**Figure S2d**, Supporting Information). For the XY8-based correlation spectroscopy sequence used for $^{15}N$ NV-NMR detection, we separately characterized the $T_2$ enhancement provided by the dynamical decoupling (DD) XY8-N pulse sequence.[20] The dependence of the coherence time on the number of π pulses applied during DD is shown in **Figure S3a**, Supporting Information. $T_2$ was extended to (40.01 ± 4.37) μs using 64 π pulses (**Figure S3c,** Supporting Information), comparable to previously reported values for shallow NV centers.[20,29]

In conventional NV-based NMR, the oscillating field from precessing nuclear spins is detected with DD sequences such as XY8-N, or Hahn-echo.[8,9] These sequences suppress low-frequency noise and act as a narrow-band filter centered at the nuclear Larmor frequency.[8,9] Their spectral resolution, however, is limited by the NV coherence time ($\approx 1/T_2$). DD sensing therefore works best for long-$T_2$ NV centers[30] and for targets precessing at MHz frequencies. For strongly coupled nuclei, such as proximal $^{13}C$ with large hyperfine interactions, a simple Hahn echo is sufficient.[31,32] Here we aim to detect $^{15}N$ nuclear spins in a $KNO_3$ solution using surface NV-NMR integrated to microfluidics.[33] The nuclear spins couple to NV only through weak dipolar interactions, which requires shallow (< 10 nm) NV centers. Surface noise, however, severely shortens their $T_2$ and degrades the signal-to-noise ratio. The challenge is compounded for diluted nuclear spins in a solution at low magnetic field. The small $^{15}N$ gyromagnetic ratio (~4.316 MHz/T) and low spin density give a weak AC magnetic field in dilute aqueous solutions. To circumvent the $T_2$ limitations, we used NV spin correlation spectroscopy where the spectral resolution is limited by the NV $T_1$.[19,34] The protocol shown in **Figure 1c** consists of two identical XY8-N blocks separated by a variable inter-sequence waiting time $\tilde{\tau}$. Each block acts as DD filter with phase evolution time $2\tau$ between the $\pi$-pulses synchronized to the Larmour precession of the target nuclear spins (such as $^{15}N$, or $^{1}H$), and satisfying the condition $\tau = \frac{\tau_L}{4}$ for maximum phase pick up. A π/2 pulse maps the phase $\varphi_1$ from the first block onto the NV spin population. After $\tilde{\tau}$, the second block accumulates phase $\varphi_2$, and the NV fluorescence reads out the correlation $\langle \varphi_1 \varphi_2 \rangle$. Sweeping $\tilde{\tau}$ produces a signal oscillating at the Larmor frequency. Its decay is limited by the NV $T_1$ [19] and/or by the nuclear correlation time, which in liquids includes diffusion of the target spins through the nanoscale detection volume limits.[20]

As a first step, we tested the microfluidic-diamond device by detecting $^{1}H$ nuclear signals from DI water flowing across the diamond surface. For a close comparison with published reports, we kept the experimental conditions close to those reported previously.[21,35] Correlation spectroscopy was performed with two XY8-10 blocks at an applied magnetic field $B_{app}$ of ~43.2 mT, by implementing a train of MW $\pi$-pulses separated with time interval corresponding to the Larmor frequency of proton. The NV spin correlation signal showed a clear oscillation as a function of the waiting time $\tilde{\tau}$, persisting up to $\tilde{\tau}$ ≈20 μs (Supporting Information Section S2, **Figure S4**). The data was fitted with an exponentially damped cosine:[36]

$$S_{corr} = A \cdot \exp\left(-\tilde{\tau}/\tau_{corr}\right) \cdot co\,s(2\pi f_0 + p) + C, \qquad \textbf{(Eq.1)}$$

where $A$ is the correlation amplitude proportional to the root- mean square field generated by the nuclear spins at the NV position, $p$ is the phase, $C$ is fitting parameters, $f_0$ is the Larmor frequency, $\tau_{corr}$ is the characteristic correlation time. Fitting yields correlation time $\tau_{corr}$ = 7.19 ±2.07 μs. The Fast Fourier transform (FFT) analysis of the time-domain data shows a prominent peak centered at 1.841 MHz, corresponding to the gyromagnetic ratio of $^{1}H$ (42.577 MHz/T) at $B_{app}$ of 42.2 mT. A Lorentzian fit to this peak gives a full width at half maximum of ≈58 kHz is in-line with earlier work on proton sensing in water.[21,35] The fast decay of the correlation time is related to the molecular diffusion coefficient of water ($D_{water}$) at the diamond surface:[36]

$$D_{water} = {4d_{NV}^2}/{9\tau_{corr}}, \qquad \textbf{(Eq.2)}$$

where $d$ is the NV depth (~2.5 nm). $D_{water} = 0.382 \times 10^{-12}$ $m^2s^{-1}$. This agrees closely with the values reported for unconfined water at diamond surface.[36]

Following the detection of the $^1H$ signal from water, we turned to $^{15}N$ nuclear spins coupled to the NV center through dipolar interactions. Nanoscale $^{15}N$ NMR is intrinsically difficult. $^{15}N$ has a small gyromagnetic ratio ($\gamma$ = −4.316 MHz/T) and a magnetic moment about ten times smaller than that of $^1H$, so each spin contributes roughly 100 times less statistical-polarization signal. Its natural abundance is also only 0.37%. As a control experiment, the diamond was first cleaned using a tri-acid cleaning procedure at 493 K for four hours. NV-XY8-N spin correlation spectroscopy measurements were performed with the cleaned diamond while the microfluidic channel was empty, i.e., no $KNO_3$/DI water solution was flowing through the device. The NV-XY8-N pulse sequence was tuned to probe $^{14}N$ and $^{15}N$ nuclear spins in diamond. The diamond used in this study has a very low concentration of N (< 5 ppb) prior to ion beam implantation of $^{15}N$ ions.[26,27] As shown in **Figure S5a** (Supporting Information), no clear oscillatory signal was observed in the time-domain correlation signal. Correspondingly, the FFT of the correlation data shown in **Figure S5b** (Supporting Information) displays no $^{14}N$ peak. This is expected due to the low concentration of the most abundant isotope $^{14}N$ in our diamond (< 5 ppb).[37] Also, we did not detect any NV-NMR signal from $^{15}N$ nuclear spins in the implanted diamond (see **Figures S5c** and **S5d**, Supporting Information) despite the high (> 10 ppm) concentration of the implanted $^{15}N$ ions. This is explained by the strong hyperfine coupling (~3.15 MHz) between $^{15}N$ nuclear spins and the NV electron spins (see the inset of **Figure S2a**).[18]

Spin correlation spectroscopy was performed with XY8-1 blocks at $B_{app}$ = 42.20 mT, with the $\pi$-pulse spacing $2\tau$ set to half the $^{15}N$ Larmor period for the 10% $^{15}N$ enriched $KNO_3$ solution. **Figure 2a** shows the correlation signal as a function of $\tilde{\tau}$. It displays a Free induction decay (FID)-like oscillation that persists up to ~500 μs. A fit with **Eq.1** gives a correlation time $\tau_{corr}$ of 460 μs, about 64 times longer than for interfacial water (7.19 μs) and approaching the NV $T_1$ values (see **Figure S5**, Supporting Information). **Figures 2b-d** shows zoomed windows at different $\tilde{\tau}$ ranges, where the oscillation and the fit remain clearly resolved. **Figure 2e** shows the FFT of the correlation data. A single, well-resolved peak appears at 191.7 ± 0.1 kHz, matching the $^{15}N$ Larmor frequency at 42.2 mT. A signal-to-noise (SNR) ratio of ~14 (peak amplitude relative to the mean noise floor) is obtained. Correspondingly, **Eq.2** gives an upper bound on the effective diffusion coefficient of the $^{15}N$-bearing species $D$ of ~$6 \times 10^{-15}$ $m^2 s^{-1}$. This is at least 64 times lower than for interfacial water near the same surface. It is also about five orders of magnitude lower than the self-diffusion coefficient of nitrate in bulk aqueous $KNO_3$ at 298 K, which ranges from ~$1.93 \times 10^{-9}$ $m^2s^{-1}$ in dilute solution to ~$1.17 \times 10^{-9}$ $m^2 s^{-1}$ in concentrated (2.61 M) solution.[38] Such a low ($6 \times 10^{-15}$ $m^2s^{-1}$) $D$ indicates the formation of a nitrate-rich surface layer, discussed further below.

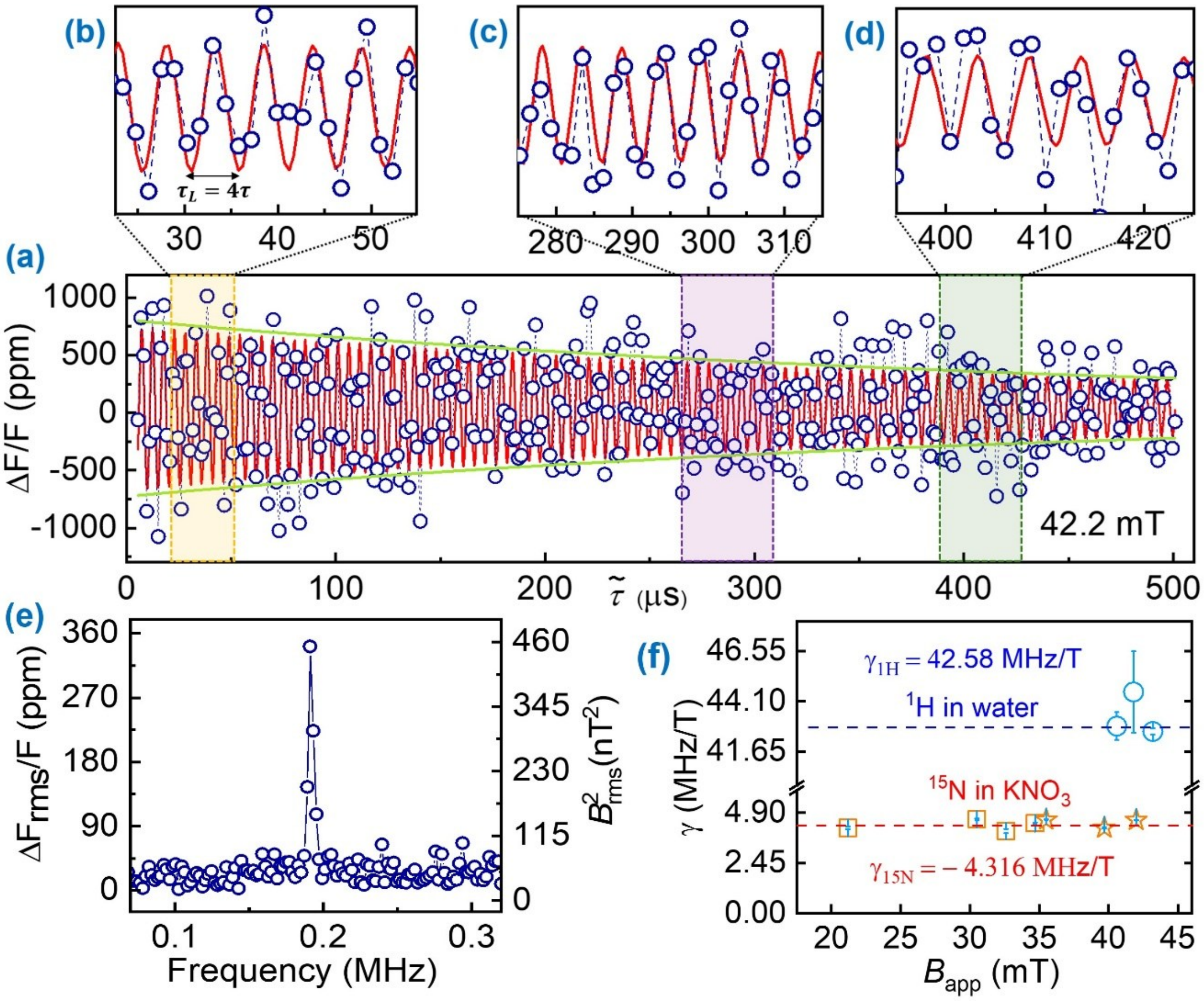

**Figure 2. Time correlation NMR signal of $^{15}$N nuclear spins in a 10% $^{15}$N -enriched $KNO_3$ solution**. **(a)** XY8-1 correlation signal (open circles) measured from a 0.2 M $KNO_3$ solution at $B_{app}$ of 42.20 mT. The solid red line represents a fit to an exponentially damped cosine function, yielding a decay constant of 462 ± 154 μs. (**b-d**) Enlarged view of the correlation curve in **(a)**, highlighting the long-lasted correlation signal. **(e)** $^{15}$N -NMR spectrum obtained from FFT of the correlation signal in **(a)** with frequency of ~0.191 MHz. **(f)** Measured nuclear gyromagnetic ratios of $^1$H in water and $^{15}$N in $KNO_3$ versus applied magnetic field amplitudes. The dashed lines correspond to the expected gyromagnetic ratio of $^1$H and $^{15}$N.

To test whether the $^{15}$N signal is consistent with an ideal 0.2 M (10% $^{15}$N) solution, we compared the measured field with that expected from statistically polarized $^{15}$N spins filling a semi-infinite homogeneous layer above an NV at depth $d$ on a (100)-oriented flat diamond surface:[39]

$$B_{RMS}^2 = \rho\left(\frac{\mu_0 \hbar \gamma_n}{4\pi}\right)^2 \left(\frac{5\pi}{96 d_{NV}^3}\right), \qquad \textbf{(Eq.3)}$$

where $\rho$ is the nuclear-spin number density and $\gamma_n$ is the nuclear gyromagnetic ratio. For the 0.2 M 10% enriched $^{15}$N $KNO_3$ solution (theoretical density of $1.204 \times 10^{22}$ spins/L), **Eq.3** gives $B_{RMS}^2$ of 10.30 nT$^2$ at $d_{NV}$ of 2.5 nm (lower limit of the NV layer) and 0.96 nT$^2$ at $d_{NV}$ of 5.5 nm (upper limit of the NV layer), respectively. These values span the implantation depth of vacancies distribution calculated from SRIM simulations in Reference 26.

We then measured the spin cerebration signals and their corresponding FFT spectra at magnetic fields in the range of 20 – 43 mT for the 10% $^{15}N$, 0.37% $^{15}N$, and $^{1}H$. See **Figure S4** (for $^{1}H$) and **Figure S7** (for 0.37% $^{15}N$) in the Supporting Information, and **Figure 2e and 3b** (for 10% $^{15}N$). **Figure 2f** shows the $^{15}N$ and $^{1}H$ Larmor presession frequency, deduced from their corresponding FFT, versus $B_{app}$. The gyromagnetic ratio is calculated individually at each magnetic field value, then plotted against the magnetic field, and compared with the theoretical values, which are represented by dashed lines in Figure 2f. This method gives a gyromagnetic ratio of 4.35 ± 0.54 $MHzT^{-1}$ for $^{15}N$ and 42.577 MHz/T for $^{1}H$, respectively, in good agreement with the theoretical gyromagnetic ratios (see the dashed lines in **Figure 2f**). In contrast, no spin correlation signal (not shown here) was observed at the $^{14}N$ due to the intrinsic nuclear quadripolar moment of $^{14}N$ (~0.751 MHz).[6] The comparison between the empty microfluidic channel control and flowing $KNO_3$ measurement provides evidence that the detected $^{15}N$ signal originates mainly from the $KNO_3$ solution rather than from the diamond or other surface defects.[33]

To calibrate the NV-NMR signal from $^{15}N$, magnetic stray field variances were obtained from a test AC field calibration performed with the XY8-1 sequence (see Supporting Information Section 3 and **Figure S8** and **Figure S9**) by using the same pulse sequence and acquisition parameters as for $^{1}H$ and $^{15}N$ nuclear spins. We converted the NV response to field variance in absolute units using the analytical correlation-signal expressions for the AC and $^{15}N$ signals.[19,40] The right axis of **Figure 2e** shows a peak field variance $B_{RMS}^2$ of 442 $nT^2$ is detected with SNR ~14. This is 43 times larger than the predicted signal for an ideal 0.2 M solution over a depth of 2.5 nm. The measured magnetic-field variance of 442 $nT^2$ corresponds to an inferred $^{15}N$ spin density of 5.16 × $10^{23}$ spins/L. This is ~43 times higher than the theoretically predicted magnetic-field variance for an ideal 0.2 M $KNO_3$ solution at an NV depth of 2.5 nm. The enhanced $^{15}N$ spin density is consistent with the accumulation of nitrate at the diamond surface and explains well the long (>4500 μs) spin correlation time, discussed further below.

We repeated the measurement on the same solution at a lower field, $B_{app}$ = 35.5 mT (**Figure 3a**). A clear oscillatory correlation signal is again observed, with $\tau_{corr}$ of ~445 μs. Its FFT shows a sharp peak at ~0.153 MHz as expected for $^{15}N$ (4.316 kHz/mT). The SNR is ~10 and the converted field variance decreased by roughly a factor of two (~195 $nT^2$). We next measured the natural-abundance (0.37% $^{15}N$) $KNO_3$ solution at the same concentration (0.2 M) and magnetic field. Its $^{15}N$ spin density (4.46 × $10^{20}$ spins/L) is 27 times lower than in the enriched solution (1.242 ×$10^{22}$ spins/L). Prior to measurements, the diamond was cleaned in a boiling tri-acid mixture to remove surface residues. To compensate for the weaker signal, we used XY8-13 blocks, which increased the phase accumulated per block, whereas XY8-1 had been sufficient for the enriched $^{15}N$ solution. **Figure 3c** shows the correlation signal recorded up to $\tilde{\tau} \approx 550$ μs at $B_{app}$ = 32.6 mT. Despite the much lower spin density, an oscillatory correlation signal is observed. Its FFT shows a peak at ~0.139 MHz with a SNR of ~3, matching the $^{15}N$ Larmor frequency. A fit with **Eq.1** gives $\tau_{corr}$ of ~420 μs, in the range of the values measured for the enriched 10% $^{15}N$ nitride solution. Interestingly the calculated magnetic field variance $B_{RMS}^2$ is ~0.45 $nT^2$, slightly higher in compression to the theoretically expected 0.38 $nT^2$ at an NV depth of 2.5 nm. This also suggest that a nitrate-rich interfacial layer that builds up over time would both slow the $^{15}N$ dynamics and raise the local spin density, explaining the longer correlation time and the excess field variance in the enriched case.

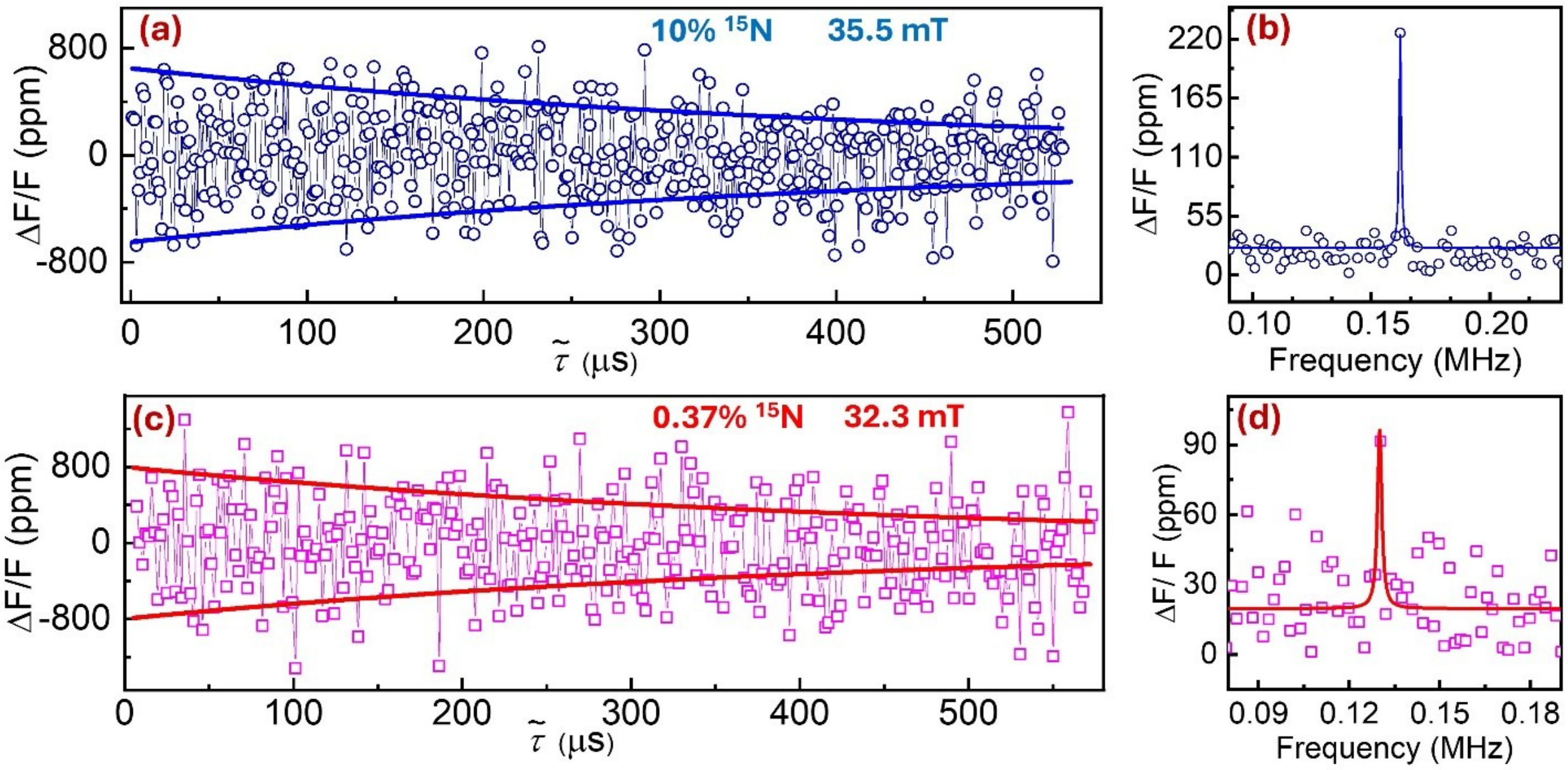

**Figure 3: Comparison of NMR correlation data for natural-abundance (0.37 %) and 10% $^{15}$N enriched $KNO_3$ solution. (a)** Correlation data for the 0.2 M, 10% $^{15}$N enriched solution recorded at $B_{app}$ of 35.5 mT. Open circles represent the experimental data, and the solid line represents the decay envelope of **Eq.1**,, yielding a correlation time of ~444.07 µs. **(b)** FFT spectrum of the data shown in (d) yielding a frequency of 0.153 MHz. **(c)** Correlation spectrum of the 0.2 M natural-abundance solution recorded at an applied magnetic field of $B_{app}$~ 32.6 mT using the XY8-13 spin-correlation sequence. The solid line represents the decay envelope of **Eq.1**, yielding a correlation time of ~450 µs. **(d)** FFT spectrum of the data in **(a)**, showing a distinct peak at frequency ~0.139 MHz.

To confirm the hypothesis of the formation/deposition of a nitrite layer, the diamond surface was analyzed by AFM and XPS before and after flowing the $KNO_3$/water solution. Prior to flowing the $KNO_3$ solution, the diamond was cleaned using a tri-acid cleaning procedure and characterized by AFM. As shown in **Figure 4a**, the cleaned diamond surface appears relatively smooth and free of visible surface contamination. The average surface roughness measured within the marked area (box) was ~0.7 nm. Just after completing the NV measurements, the diamond was removed from the microfluidic device and characterized again using AFM under the same conditions. The post-experiment AFM images in **Figure S10** (Supporting Information Section S4) for a large area (100 × 100 µm$^2$) and **Figure 4b** for a small area (10 × 10 µm$^2$) show additional surface features compared with the pristine diamond, indicating a formation of a $KNO_3$ layer during the experimental process. The average thickness of the layer is ~20 nm with a roughness of up to~ 5 nm. The AFM analysis correlates with NV measurements with the detection of a long decayed $^{15}$N spin correlation signal, influenced from the nitrite layer deposited on top of diamond.

The XPS measurements (**Figure S11a**, Supporting Information Section S4) of diamond just after following $KNO_3$ solution were used to compare the surface elemental composition and identify any changes in the chemical species present at the diamond surface.[26] The XPS results from the diamond with $KNO_3$ for C 1s, K 2p, and O 1s are plotted in **Figures S11b**, **S11c**, and **S11d**, respectively. In addition to the expected carbon C 1s and nitrogen N 1s peaks, the presence of K and O further confirms the presence of a residual $KNO_3$ film. The binding energies for K $2p_{3/2}$, $2p_{5/2}$, and O 1s are 292.98, 295.88, and 532.28 eV, similar to the values reported in literature.[41] The binning energy of N 1s peak for diamond with $KNO_3$ is ~399.4 eV (see **Figure 4c**), slightly below the measured values of 406.8 – 407.2 eV in $KNO_3$ powder.[41] The binning energy of N 1s peak for the diamond without $KNO_3$ is ~399 eV, comparable to what was measured in the lietarture.[42] We see a significant increase of the XPS intensity of N 1s peak of diamond/$KNO_3$ by ~5 times (from the base line) in comparison to the one measured on just nitrogen implanted diamond (see **Figure 4c**). As discussed above, the implanted nitrogen (15N) does not contribute to the NV-NMR 15N signal coming from KNO3 due to the ~3.15 MHz hyperfine coupling. This observation signifies the presence of a nitrite layer as confirmed by AFM and NV-NMR measurements of $^{15}N$ nuclear spins, discussed above.

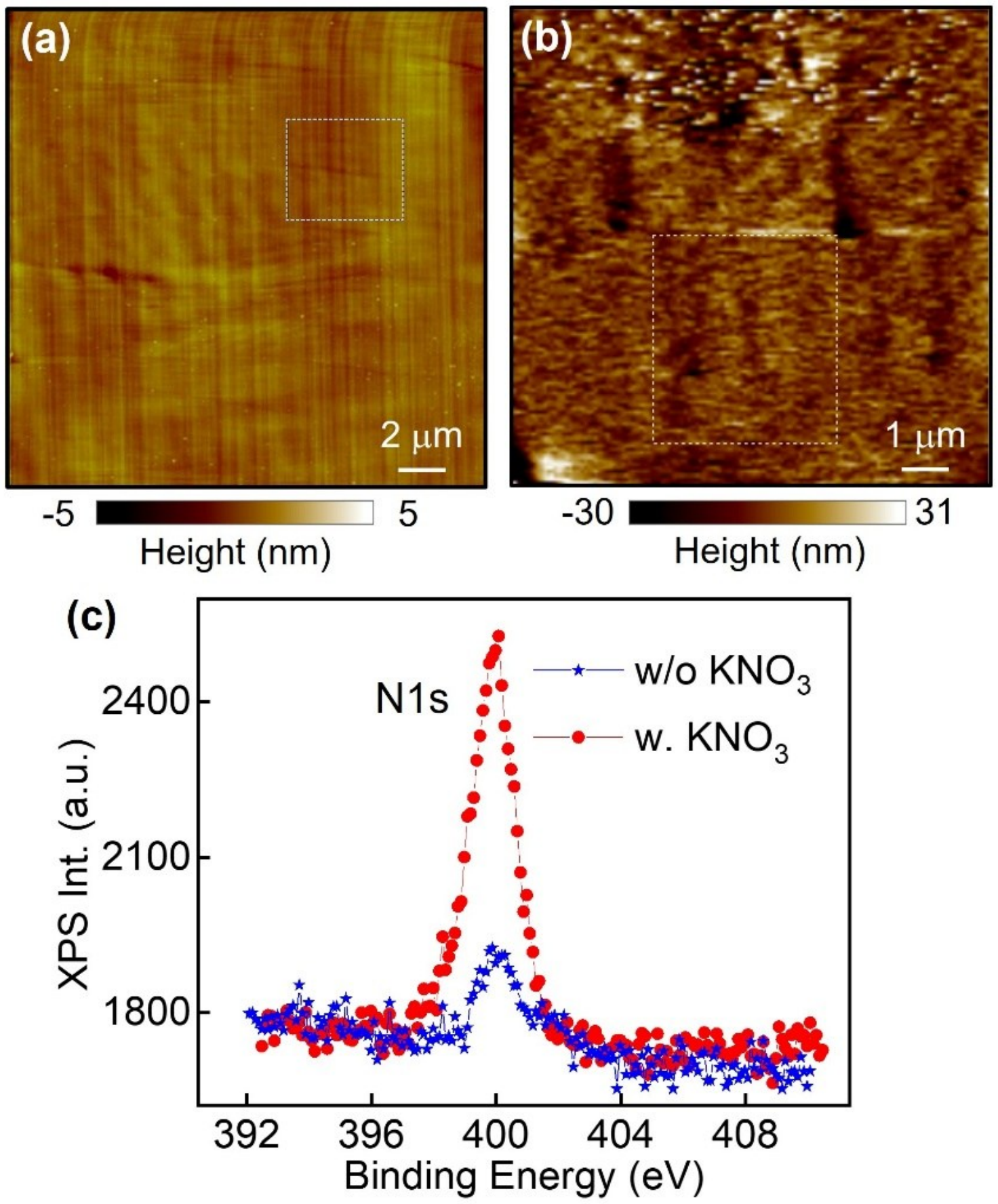

**Figure 4:** Topographic images of the diamond surface (a) before and (b) after flowing the $KNO_3$ solution over it and allowing it to dry. The roughness analysis was performed in the dashed squares. (c) Measured XPS spectrum of N1s electrons without (scattered stars) and with (scattered dots) $KNO_3$ confirming the presence of nitrogen at the surface of the diamond.

To conclude, by using microfluidic diamond quantum sensing platform based on NV-NMR we detected $^{15}$N spin correlations in naturally abundant (0.37%), low (0.2 M) concentrated $KNO_3$ solution at room temperature and at moderate magnetic fields (20 – 43 mT). A gyromagnetic ratio of 4.35 ± 0.54 MHz/T was deduced from field dependence measurements of the NV XY8-13 spin correlation, in good agreement with the theoretical value of $^{15}$N. The $^{15}$N correlation decays within ~500 μs, mainly limited by the NV $T_1$ and not by the molecular self-diffusion of the $KNO_3$ molecules in water. The 10% enriched $^{15}$N nitrite solution produces a strong stray field with $B_{RMS}^2$ of ~442 nT², good SNR of ~14, and long correlation time (> 450 μs), suggesting the deposition of a nitrite layer on the diamond surface. Surface analysis by AFM and XPS of the diamond surface after flowing the $KNO_3$ solution, confirms the presence of ~10–20 nm layer of $KNO_3$ with increased concentration of nitrogen. Further NV-NMR measurements on 10% $^{15}$N-enriched $KNO_3$ solution reveal an increased SNR signal with shorter averaging times.

These results highlight the sensitivity and local probe nature of NV quantum sensors to surface analytes, relevant to chemical analysis.[16,28] Future work with controlled surface treatments, different concentrations and temperatures will help clarify how such surface layers form and how they affect molecular motion.

## ASSOCIATED CONTENT

### Data Availability Statement

The data that support the findings of this study are available from the corresponding author upon reasonable request.

## AUTHOR INFORMATION

### Author Contributions

A.L. conceived the idea of the project, designed the experiments, and supervised the project. S.I. performed the NV, XPS, and AFM measurements with assistance of J.K and R.T. S.I made the microfluidic diamond device and analyzed the data with assistance of J.K. S.L. and S.-H.L performed AFM and XPS, and helped in making the microwave loops and microfluidic device. S.I, J. K., and A.L. wrote the manuscript with contributions and feedback from all authors.

### Notes

The authors declare no competing financial interest.

## ACKNOWLEDGMENTS

A.L. acknowledge the National Science Foundation (NSF) through Award 2426522 and the University of Nebraska-Lincoln (UNL) Grand Challenges catalyst award entitled “Quantum Approaches addressing Global Threats”. S.-Y.L thanks support from NSF through the EPSCoR RII Track-1 program under Award OIA-2044049. The research done at the University of Nebraska-Lincoln was performed in part in the Nebraska Nanoscale Facility: National Nanotechnology Coordinated Infrastructure and the Nebraska Center for Materials and Nanoscience (and/or NERCF), supported by NSF Award 2025298. We thank Victor Acosta and Janis Smits from the University of New Mexico for their support during the setting of the NV-NMR microscope and NV sensitivity estimation.

# Supporting Information

## Section S1. Nitrogen-Vacancy Nuclear Magnetic Resonance (NV-NMR) Spectroscopy

***Microfluidic NV-NMR setup:*** **Figure S1a** shows the nitrogen vacancy (NV) doped diamond attached to a glass coverslip patterned with two microwave (MW) antenna made of Au/Ta or Cu/Ta.[43,27,26,44] The diamond is cured to glass using UV glue (see Methods in the main texts). **Figure S1b** displays a photograph of the fully assembled microfluidic device, showing the inlet and outlet ports used for continuous circulation of the aqueous $KNO_3$ solution. A programmable double syringe pump (World Precision Instrument) was used to maintain a controlled and continuous flow of the potassium nitrite $KNO_3$/DI water solution across the NV sensing region. **Figure S1c** shows the diamond microfluidic NV-NMR setup used to detect $^{15}N$ nuclear spins in $KNO_3$/DI water solution. The microfluidic device is mounted on a sample stage positioned above an oil objective (NA of 1.25), which focuses the 532 nm green laser onto the diamond and collects the NV-center fluorescence in the 650–750 nm range. The laser power is ~180 mW focused on ~32 µm × 32 µm, corresponding to a laser intensity of ~0.22 mW/µm$^2$. The fluorescence is focused on avalanche photon detector (Thorlabs APD410A), connected to a Yokogawa oscilloscope (DL9040L) for NV-NMR acquisition. A permanent magnet (K&J Magnetics) is aligned along the NV quantization axis [111] for (100) diamond and used to NV-NMR measurements in the range of 20 – 50 mT. The motion of the microfluidic device is controlled by $X$, $Y$, and $Z$ piezoelectric motors for precise laser focus and NV/diamond positioning. Further details about the NV (optical, electronic, and MW) setup can be found in References [27,43].

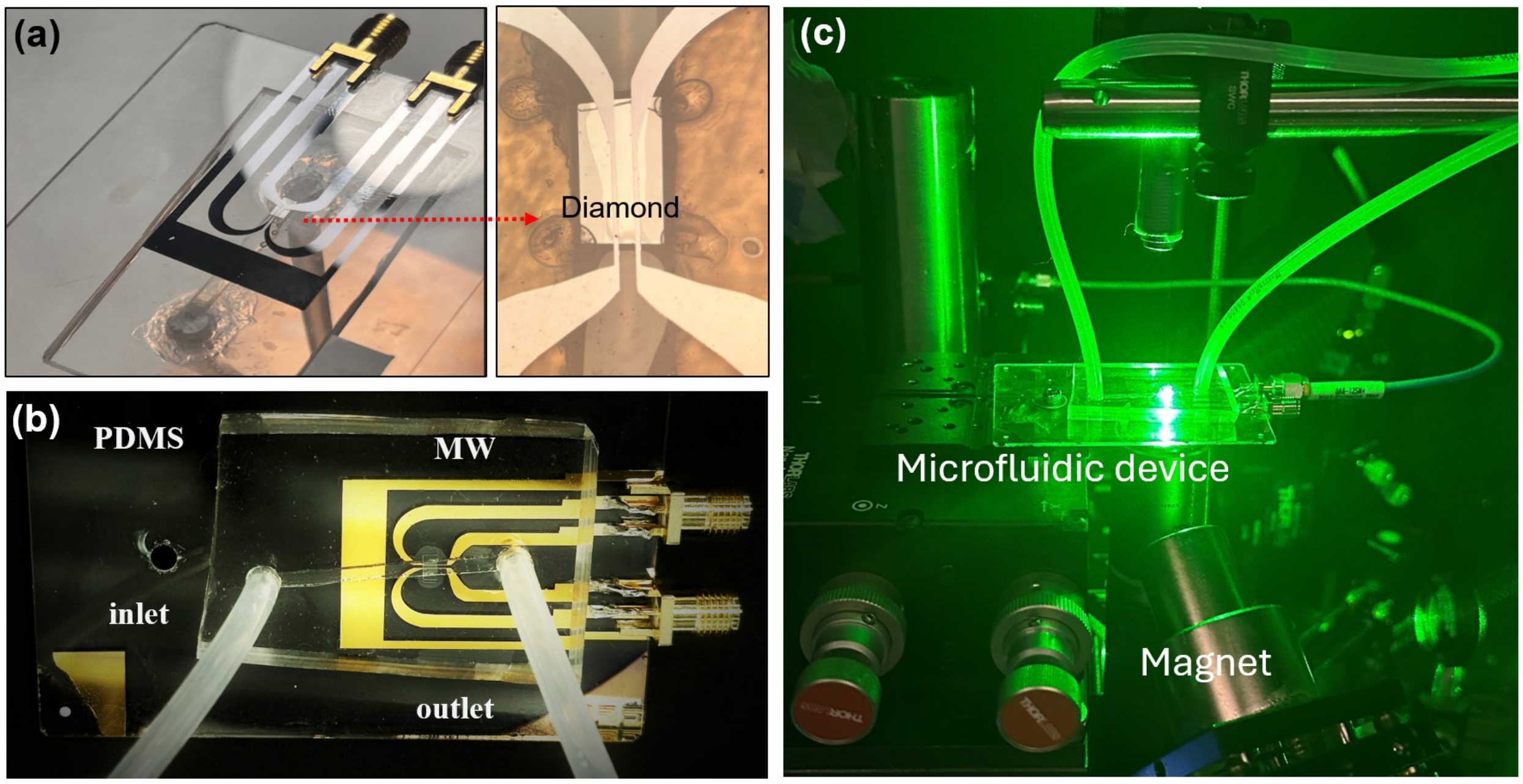

**Figure S1:** (a) Left: Photograph of the glass PCB patterned MW transmission lines, with diamond mounted on it. Right: A zoomed optical image of the diamond bonded to the microwave transmission line using UV-curable adhesive. (b) Fully assembled microfluidic device integrated with the diamond sensor and connected to inlet and outlet tubing for continuous flow of the $KNO_3$/DI water solution. (c) A photograph of the room-temperature microfluidic NV-NMR experimental setup used for $^{15}N$ sensing.

***NV ensemble characterization:*** **Figure S2a** shows representative optically detected magnetic resonance (ODMR) spectrum at $B_{app}$ of 40.96 mT at MW power of 0.32 W. A resonance frequency of 4.016 GHz was obtained for the subsequent coherent NV spin measurements at $B_{app}$ of 40.96 mT. At weak MW power (3 mW), two peaks appear in the ODMR spectrum with a splitting of ~3.15 MHz, associated with the $^{15}$N nuclear spin hyperfine coupled to the NV electronic spin.[18,45] **Figure S2b** illustrates the NV Rabi oscillation measurements at $B_{app}$ of 40.96 mT to determine π and π/2 for $T_1$ (spin lattice relaxation time) and $T_2$ (spin-spin relaxation time) measurements. The observed oscillations result from coherent MW-driven rotations of the NV electronic spin between $|0\rangle$ and $|+1\rangle$ spin states. The Rabi frequency is 21.71 MHz, corresponding to π/2 of ~12 ns and π of 23 ns. The Rabi decay is 0.768 ± 0.2 µs, that is expected from shallow NV ensembles.[26,27] We next measured NV $T_1$ of bare diamond (no $KNO_3$ or water) and found $T_1$ of 1.499 ± 0.11 ms (**Figure S2c**). Hahn echo measurements in **Figure S2d** yield $T_2$ of 4.23 ± 0.07 µs. The measured signal shows the characteristic collapse-and-revival behavior arising from the interaction of the NV electronic spin with the surrounding naturally abundant $^{13}$C nuclear-spin bath in the diamond lattice,[19] in addition to the dark surface spins.[46,47]

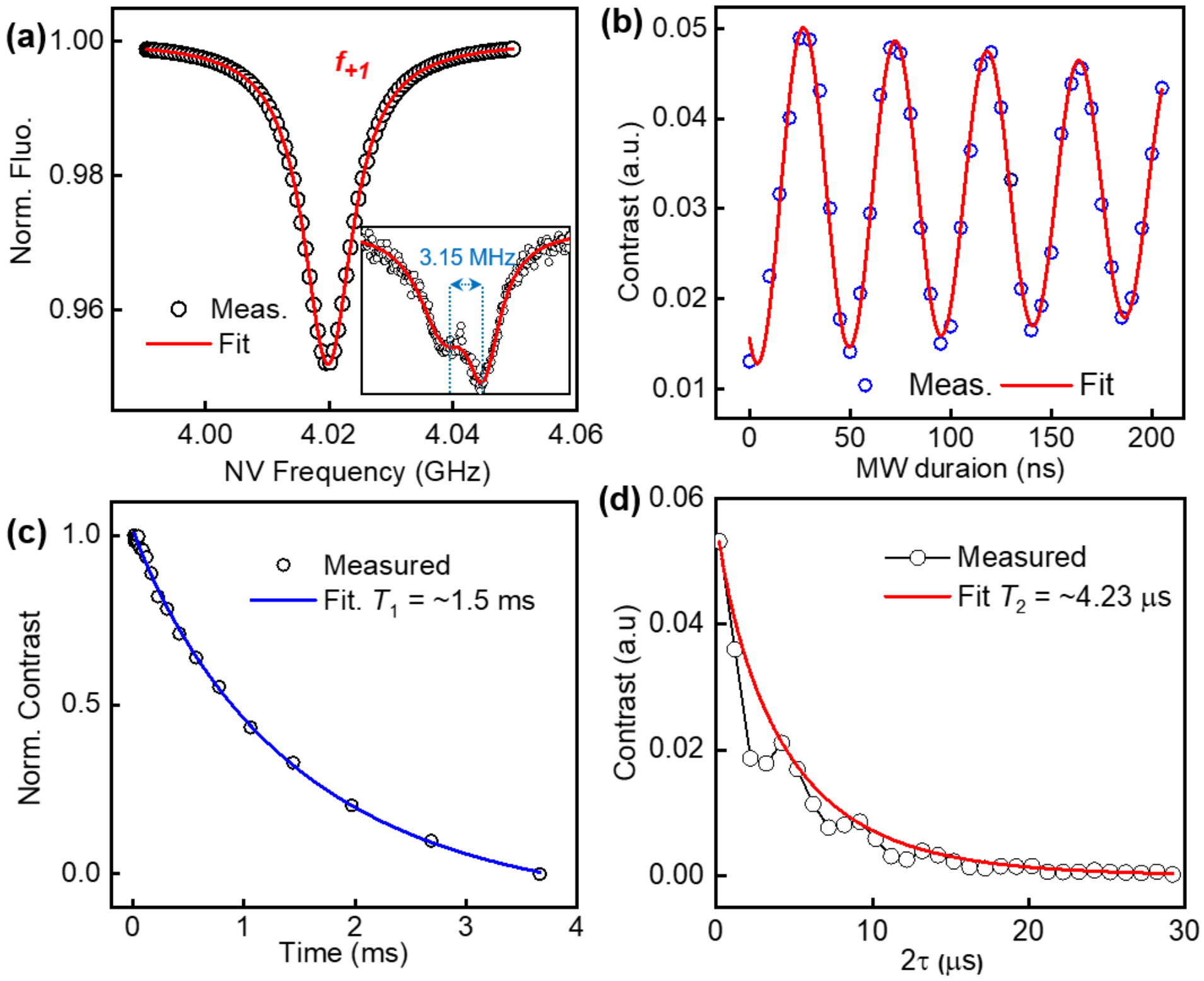

**Figure S2:** (a) Measured (open circles) and Lorentzian fitted (solid line) continuous-wave (CW) ODMR spectrum for $f_+$ ($|0\rangle \leftrightarrow |+1\rangle$) NV spin transition, recorded at $B_{app}$ of 40.96 mT and MW power of ~0.32 W. Inset of (a): Measured (open circles) and double Lorentzian fitted (solid line) ODMR spectrum taken at a MW power of 3 mW, showing the hyperfine coupled $^{15}$N nuclear spins of the NV spins. (b) Measured (open circles) NV Rabi oscillations at MW frequency of 4.016 GHz, corresponding to the ODMR resonance in (a). The red solid line shows the fit to an exponentially decaying oscillatory function, yielding a Rabi frequency of 21.71 MHz and decay of 0.768 ± 0.2 ms. (c) Measured (open circles) NV spin lattice relaxation time ($T_1$) of the bare diamond. The solid blue line represents the single-exponential fit to the experimental data, yielding $T_1$ of 1.499 ± 0.11 ms. (d) Measured (open circles) and exponential fitted (solid line) NV Hahn echo revivals of $^{13}$C at 40.96 mT, yielding a $T_2$ of 4.23 ± 0.07 µs.

***Dynamic decoupling XY8-N pulse sequence:*** To increase the NV sensitivity to external surface nuclear spins, XY8-N sequence (**Figure S3a**) was used to improve the NV coherence time $T_2$.[20] **Figure S3b** displays $T_2$ (fluorescence intensity vs total evolution time) curves for N = 2, 4, 6, and 8, respectively. By increasing the number of π pulses to 64, $T_2$ is extended to 40.01 ± 4.37 μs. As expected, $T_2$ increases monotonically with the number of π pulses (**Figure S3c**), indicating that increasing the number of pulses effectively improves the NV coherence time.[20]

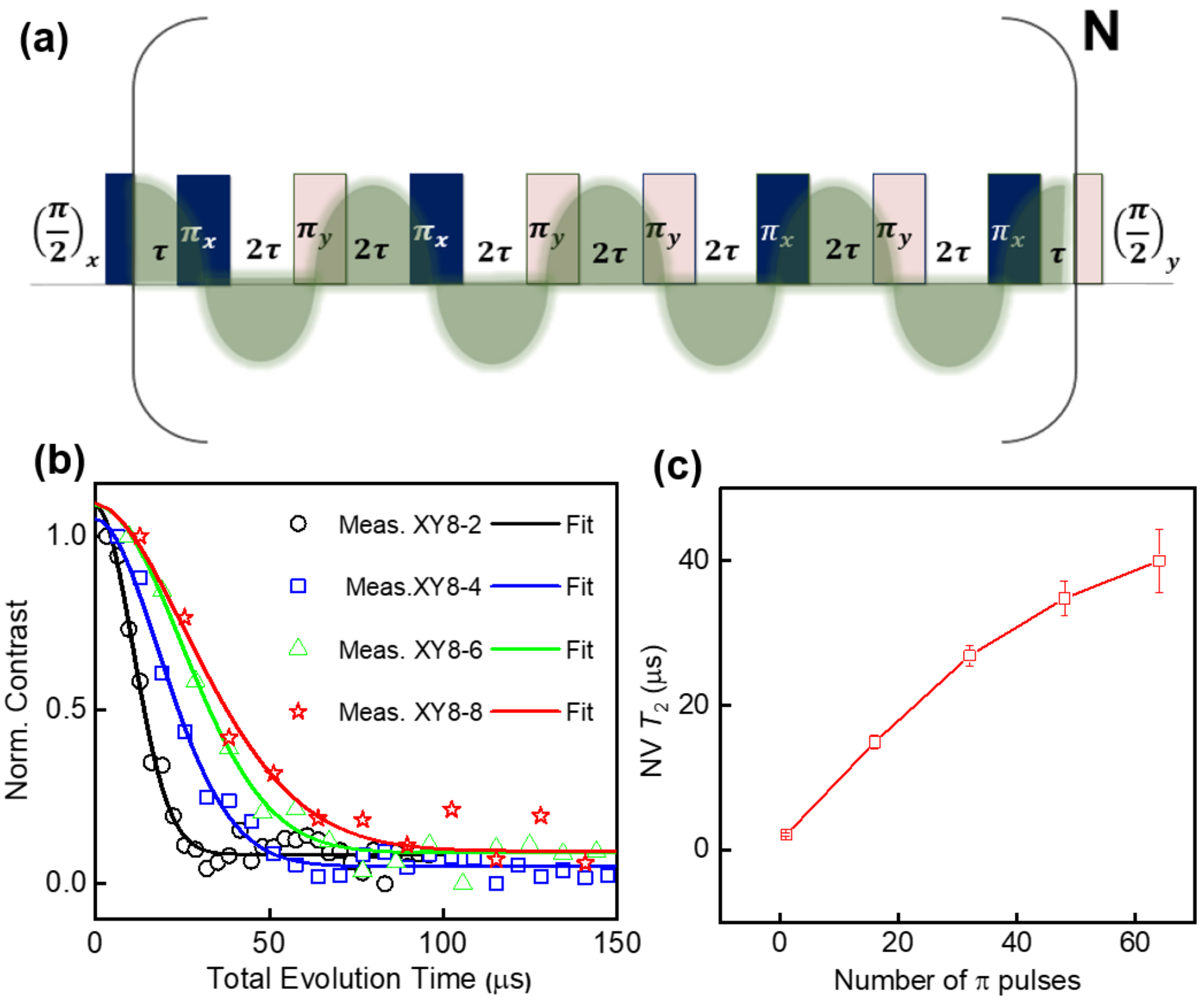

**Figure S3:** (a) A schematic of the dynamical decoupling XY8-N pulse sequence. (b) Measured normalized (scattered) and exponential-fitted (solid lines) fluorescence intensity versus total evolution time for N = 2, 4, 6, and 8 of $f_+$ spin transition at $B_{app}$ of 41.4 mT. The fitting yields to $T_2$ of 14.91 ± 0.93 μs, 26.88 ± 1.48 μs, 34.79 ± 2.43 μs, and 40.0 ± 4.37 μs for N = 2, 4, 6, and 8, respectively. (c) $T_2$ vs number of π pulses, showing an efficient dynamic decoupling from NV environment noise that led to an increase of $T_2$.

### Section S2. NV-NMR measurements on water and $KNO_3$ solutions

***NV-NMR of $^1$H in water:*** We first tested the microfluidic device by detecting the protons ($^1$H) NMR signal from DI water continuously flowing through the microfluidic channel in **Figure S1b**. The $^1$H NMR signal was measured using spin-correlation XY8-N (N=10) dynamical decoupling protocol,[20] with $4\tau = 1/f$, $f$ is the proton Larmor frequency. At $B_{app}$ of 42.2 mT, a clear oscillatory signal (Figure S4a) with a period of ~0.58 μs. As shown in **Figure S4b**, this signals decays sharply with a decay time of 7.19 ± 2.07 μs, which is consistent with previous measurements.[20,21] To further analyze the signal in the frequency domain, the fast Fourier transform (FFT) was performed on the time-domain correlation data. The resulting spectrum in **Figure S4c** shows a clear peak centered at~1.841 MHz, consistent with the expected proton Larmor frequency at 42.2 mT with a gyromagnetic ratio of 42.58 MHz/T. The peak was fitted using a Lorentzian profile, giving a linewidth of approximately 58 kHz.[21] These results confirm that our microfluidic diamond device can reliably detect proton NMR signals under continuous-flow conditions.

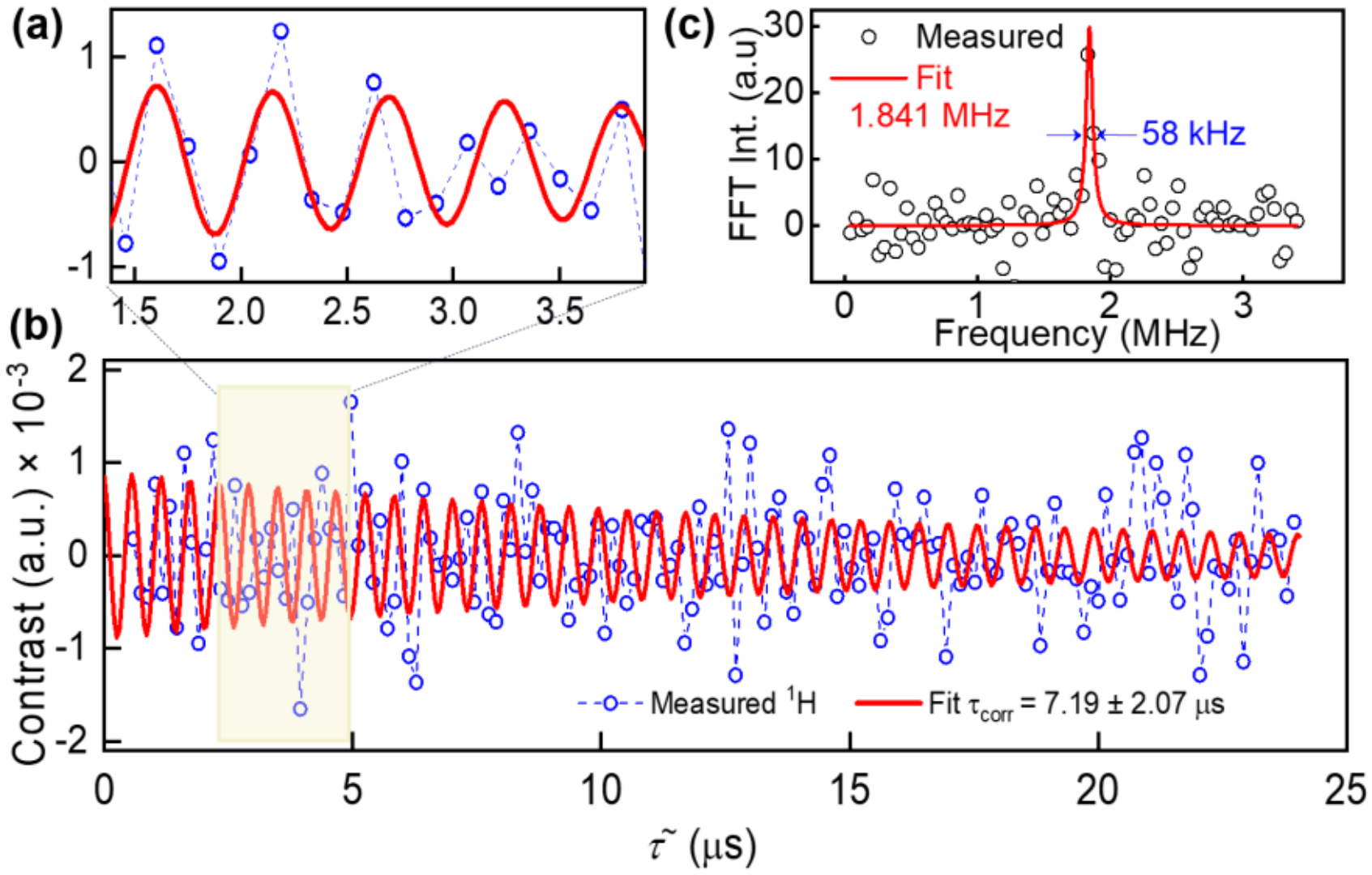

**Figure S4:** (a) NV-XY8-10 correlation signal (open circles) recorded from $^1$H nuclei in DI water flowing over the diamond surface at an external magnetic field of ~42.2 mT. The solid red line is the best fit to an exponentially damped oscillation, corresponding to a correlation decay time around 7 μs. (b) Zoomed-in view of the correlation signal over the 2–5 μs interval, illustrating the quality of the fit to the experimental data. (c) FFT spectrum of the XY8-10 correlation signal. The solid red line is a Lorentzian fit.

**Figure S5** shows the NV spin correlation spectroscopy measurements with the corresponding FFTs for $^{14}$N and $^{15}$N nuclear spins with cleaned diamond, i.e., no $KNO_3$/DI water solution was flowing through the device. No signal is detected, as discussed in detail in the main text.

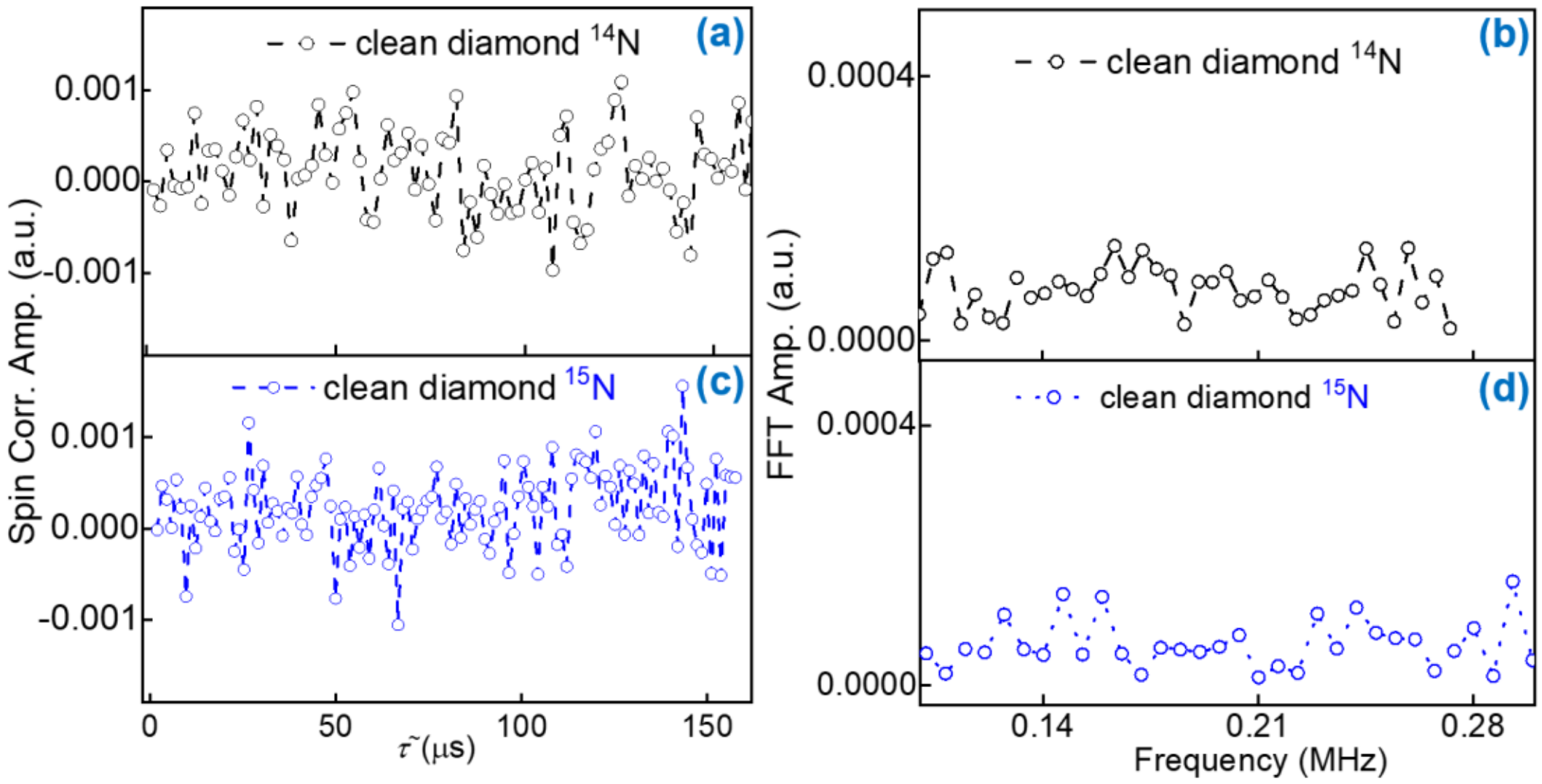

**Figure S5:** (a, c) NV-XY8-13 spin correlation measurements of the NMR signals from $^{14}$N and $^{15}$N in diamond, respectively, recorded at $B_{app}$ of ~42 mT in the absence of $KNO_3$. (b, d) Corresponding FFT spectra of the correlation signals shown in (a, c) respectively.

***NV $T_1$ relaxometry in the presence of $KNO_3$ solution:*** The nuclear spins correlation time, $T_{corr}$, is $\leq$ the NV $T_1$,[19] providing an increased sensitivity of NVs to detect external individual protons.[48] The phase accumulated by the NV spins during the first XY8 block is mapped onto the longitudinal spin population and preserved during the evolution period. The stored population is subsequently converted back into a measurable phase during the second XY8 block, enabling correlation of the phases accumulated in the two blocks.[19] Thus, the $T_1$ time determines the available storage window for the accumulated phase and is an important consideration when selecting the evolution delay for the spin correlation measurements of $^{15}N$ nuclear spins in $KNO_3$. Prior to measuring the spin correlation, we measured $T_1$ for applied magnetic fields in the range of 25 – 42.2 mT, as shown in **Figure S6(a–d)**. The measured curves were fitted with a single-exponential decay function to extract the corresponding $T_1$ values in the range of 0.49–0.8 ms. The decrease of $T_1$ for certain values of magnetic field or concentrations of $KNO_3$ can be explained by the presence of paramagnetic nitrogen spins[26,46] or surface spins.[19,49]

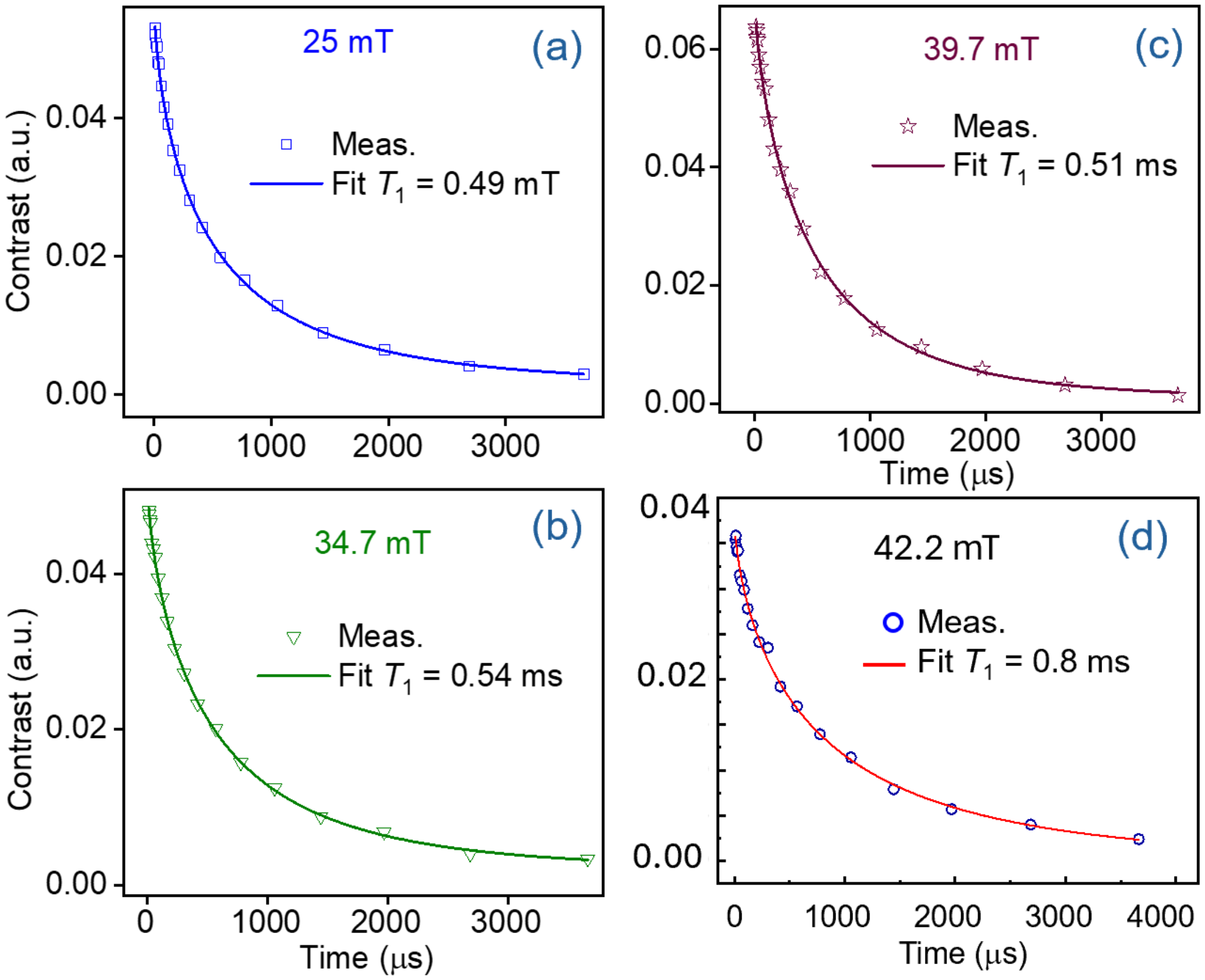

**Figure S6.** Measured (scattered) and fitted (solid line) NV $T_1$ (fluorescence vs time) curves in the presence of $KNO_3$ solution at different magnetic fields of 25 mT (a), 34.7 mT (b), 39.7 mT (c), and 42.2 mT (d).

***NV-NMR of $^{15}N$ in $KNO_3$ solution:*** To check whether the measured signal comes from $^{15}N$ spins, we measured the spin cerebration signals (**Figure S7a**) and their corresponding FFT spectra (Figure S7b) at magnetic field ($B_{app}$) in the range of 21.2 – 34.7 mT for the naturally abundant (0.37% of $^{15}N$ $KNO_3$ solution). As discussed in the main text, the linearly fitted curve gives a gyromagnetic ratio of 4.35 ± 0.54 MHzT$^{-1}$, in good agreement with the theoretical gyromagnetic ratio of $^{15}$N.

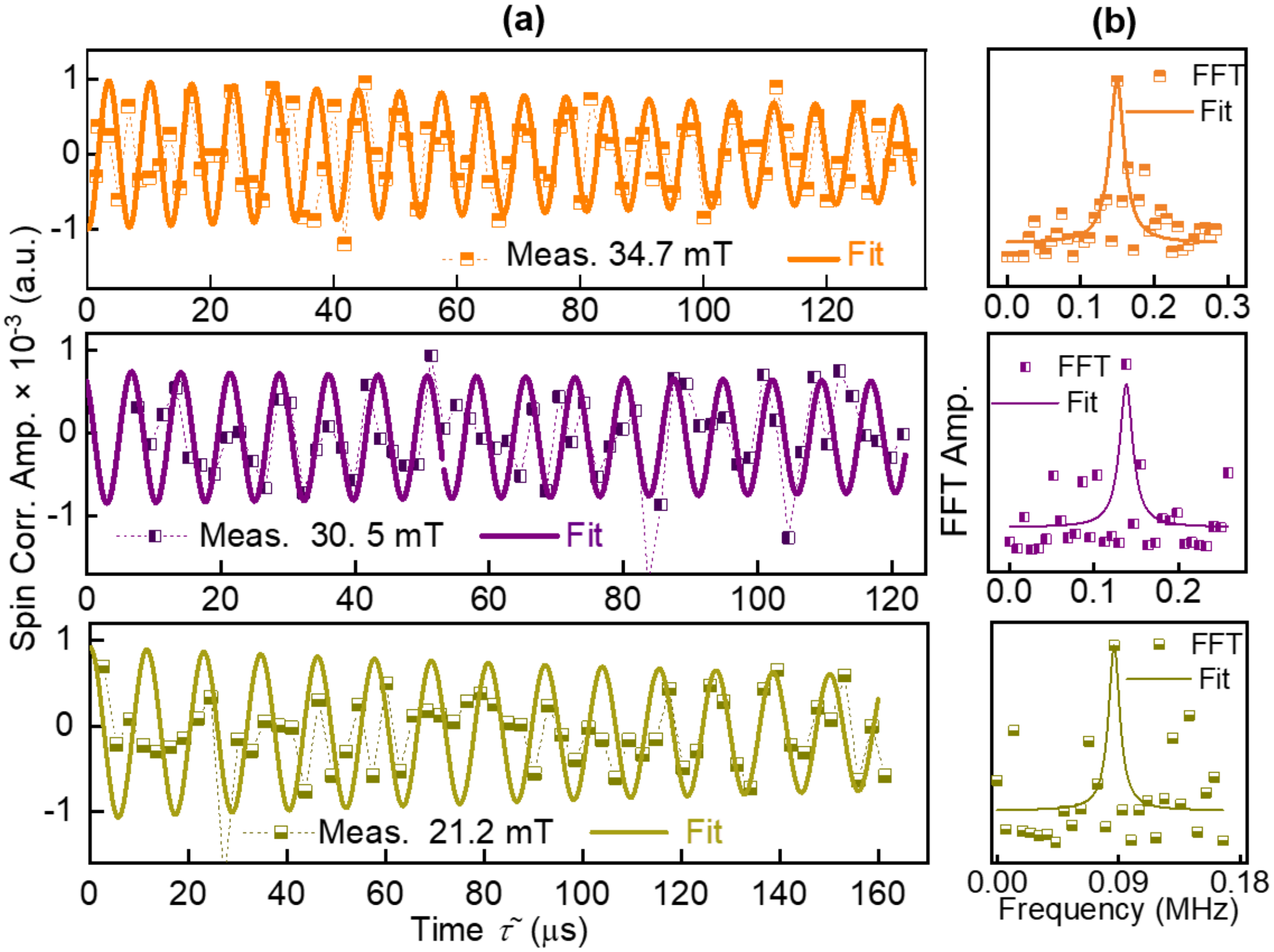

**Figure S7:** (a) NV-NMR, XY8-13 correlation spectroscopy data acquired from an aqueous $KNO_3$ solution containing naturally abundant $^{15}$N in different external magnetic fields. Solid curves represent exponential oscillatory fits to the experimental data. (b) FFT spectra corresponding to the correlation data shown in (a). Solid curves represent Lorentzian fits used to determine the Larmor frequencies of $^{15}$N nuclear spins.

## Section S3. AC calibration and NV sensitivity

To calibrate our microfluidic quantum sending device, we applied a known external AC signal in the absence of any solution (water or $KNO_3$) driven by a function generator. The excitation frequency was set at the expected $^1$H and $^{15}$N Larmor precession frequencies at a magnetic field of 42.2 mT. Using the spin-correlation XY8-$N$ pulse sequence, we successfully detected the applied AC signals at both frequencies. The corresponding time-domain correlation signals for $^1$H (N = 10) and $^{15}$N ($N$=1) are shown in **Figures S8a** and **S8c**, respectively, while their corresponding FFT spectra are displayed in Figures S8b and S8d. The FFT spectra exhibit distinct peaks at ~1.7 MHz and ~0.18 MHz, consistent with the expected $^1$H and $^{15}$N Larmor frequencies at $B_{app}$ of 42.2 mT. This calibration demonstrates the functionality of our microfluidic NV-NMR device to detect AC magnetic signals relevant to the detection of nuclear spins in nitrites.

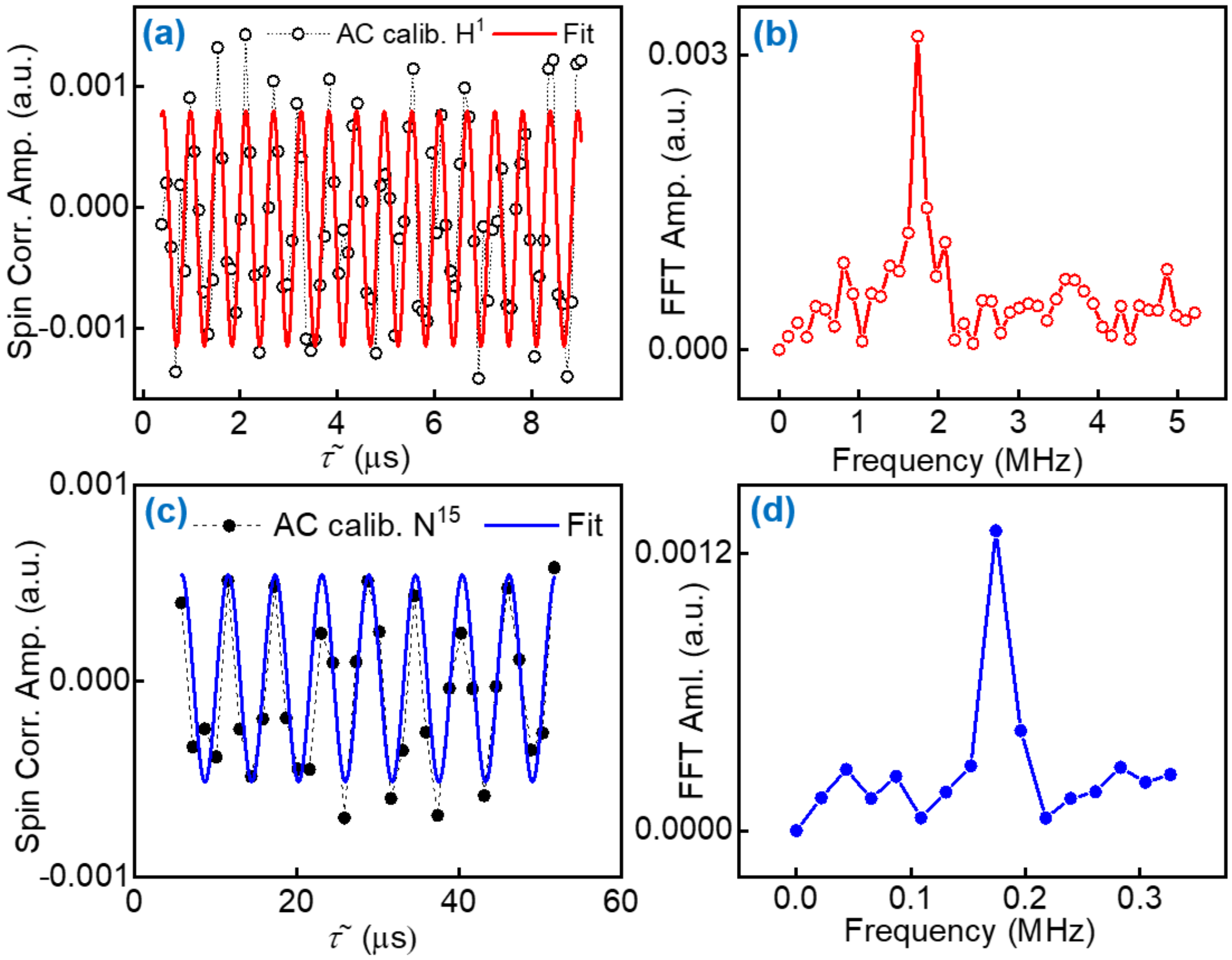

**Figure S8:** Measured (scattered) and fitted (solid line) XY8N spin correlation acquired for AC fields corresponding to $^{1}H$ (a, N = 10) and $^{15}N$ (b, N = 1) frequencies, respectively. The red and blue solid lines represent sinusoidal fits to the corresponding correlation signals. Corresponding FFT spectra for AC fields at $^{1}H$ (a) and $^{15}N$ (b) frequencies, respectively.

To calibrate the magnetic-field sensitivity of the microfluidic NV-NMR device, we adopted the calibration method described in detail in Ref [50]. In this approach, a known RF test signal is applied to the transmission line integrated with the microfluidic device, and the resulting NV fluorescence response is measured using an XY8-$N$ spin-correlation pulse sequence. The amplitude of the applied RF test signal, $V_{\text{test}}$, is systematically varied while maintaining the test frequency, $f_{\text{test}}$, at the target nuclear Larmor frequency. Here, $V_{\text{test}}$ denotes the peak-to-peak voltage applied to the transmission line at $f_{\text{test}}$.

Following the method described in Ref. [50], the NV test-signal response as a function of the applied RF voltage can be expressed as:

$$V_{NV} = V_{max} \left| \sin\left(4\sqrt{\frac{2}{3}} * \gamma_{nv} * \kappa * V_{test} * \tau_{tot}\right) \right|, \qquad \textbf{Eq.S1}$$

where $V_{\max}$ represents the maximum NV response and $\tau_{\text{tot}}$ is the total NV phase-accumulation time, $\gamma_{NV}$ is the gyromagnetic ratio of the NV center. The parameter $\kappa = \frac{B_{rms}}{V_{rms}}$ is the voltage-to-magnetic-field scaling factor, which provides the calibration between the applied RF voltage and the corresponding magnetic-field amplitude $B_{\text{test}}$ generated at the NV sensing volume. The factor $\sqrt{(2/3)}$ accounts for the rms-to-peak conversion of the test field and its projection onto the NV axis for a (100) diamond surface.[20]

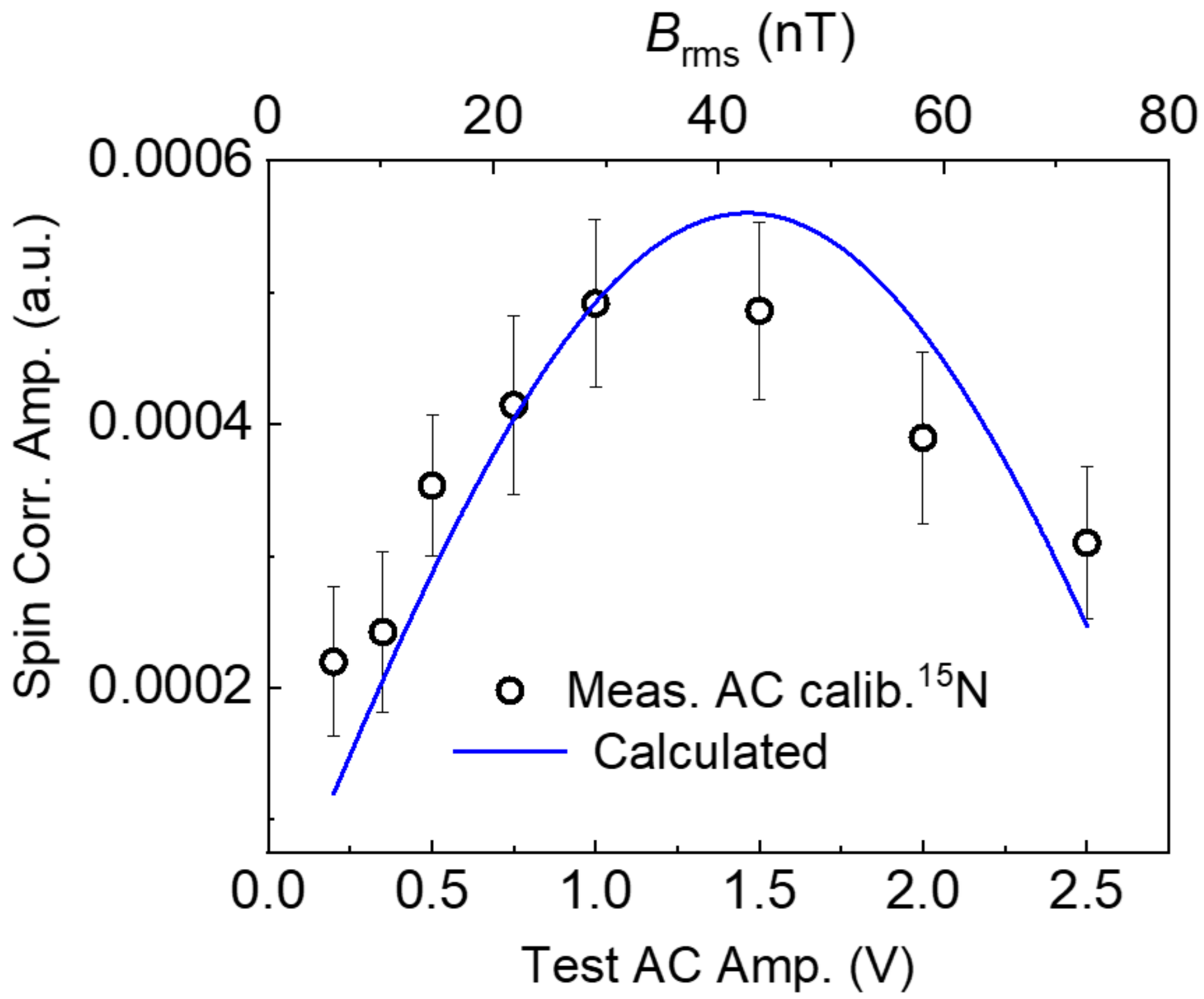

**Figure S9: Magnetic field Sensitivity of the microfluidic device for $^{15}$N Larmor frequencies.** NV fluorescence response as a function of the applied RF voltage to the microfluidic device. An RF test field at 0.18 MHz, corresponding to the $^{15}$N Larmor frequency, was applied to the device. The NV fluorescence response was measured using an XY8 spin-correlation pulse sequence with $\tau$ = 1.39 µs to detect the *z* projection of the NV center while varying the applied RF voltage from 0.2 to 2.5 V, corresponding to an estimated $B_{rms}$, plotted on the top *x*-axis.

To experimentally determine this scaling factor, we measured the NV fluorescence response with systematically varying $V_{pp}$ amplitude. The corresponding calibration measurements were performed at nitrogen $^{15}$N Larmor frequencies at 42.2 mT. The experimental data were fitted using equation **Eq.S1** as shown in **Figure S9**, and the resulting fit parameters were used to determine the corresponding magnetic-field scaling factor $\kappa$= 81.66 ± 3.74 nT/$V_{pp}$ for $^{15}$N. This calibration establishes the relationship between the applied RF voltage and the magnetic field experienced by the NV centers, as shown in **Figure S9** top x-axis, and is subsequently used to quantify the magnetic-field sensitivity of the microfluidic device.

### Section S4. Surface Analysis of diamond without and with $KNO_3$

An atomic force microscope (AFM) Innova from Bruker was used to determine the roughness of the diamond prior and after the flowing $KNO_3$/DI water solution. **Figure S10** shows a large diamond area (100 × 100 µm) after flowing $KNO_3$ and letting it to dry. Further discussion is detailed in the main text.

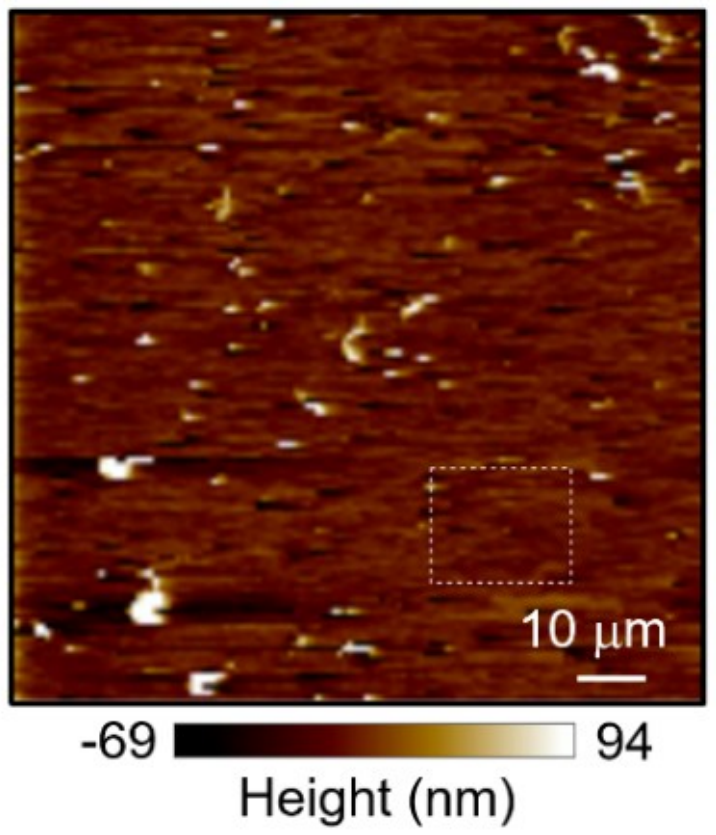

**Figure S10:** AFM image of diamond after flowing $KNO_3$ solution and letting it dry of a 100 µm ×100 µm region. The roughness analysis was performed in the dashed square.

To further confirm the formation of $KNO_3$ layer on the diamond surface, XPS was performed before and after flowing $KNO_3$ solution. XPS examination of nitrogen content in diamond prior and after flowing $KNO_3$/DI water solution was performed using Thermo Scientific Al K-Alpha XPS system under an ultrahigh vacuum pressure of $1 \times 10^{-9}$ mbar.[51] The X-ray beam had a pass energy of 50 eV and a diameter of 400 µm. The number of averaged sweeps for each element (N, O, C, and K) was adjusted to optimize the signal to noise ratio, typically ranging from 20 to 100 sweeps (see **Figure S11**). The XPS data acquisition and analysis was performed using the Avantage™ software package.[26] For the data analysis, we calibrated the XPS spectra relative to the first carbon C 1s peak in **Figure S11a** at a binding energy of 284.8 eV.[26] In our diamond, two maxima were observed in **Figure S11b**, one at ~284 eV and the second one at~285 eV, in a good agreement with previous XPS measurements on implanted electronic grade diamond.[49] Figures **S11a** and **S11d** show the XPS peaks for potassium (K $2p_{3/2}$, $2p_{1/2}$) and oxygen (O 1s), respectively.

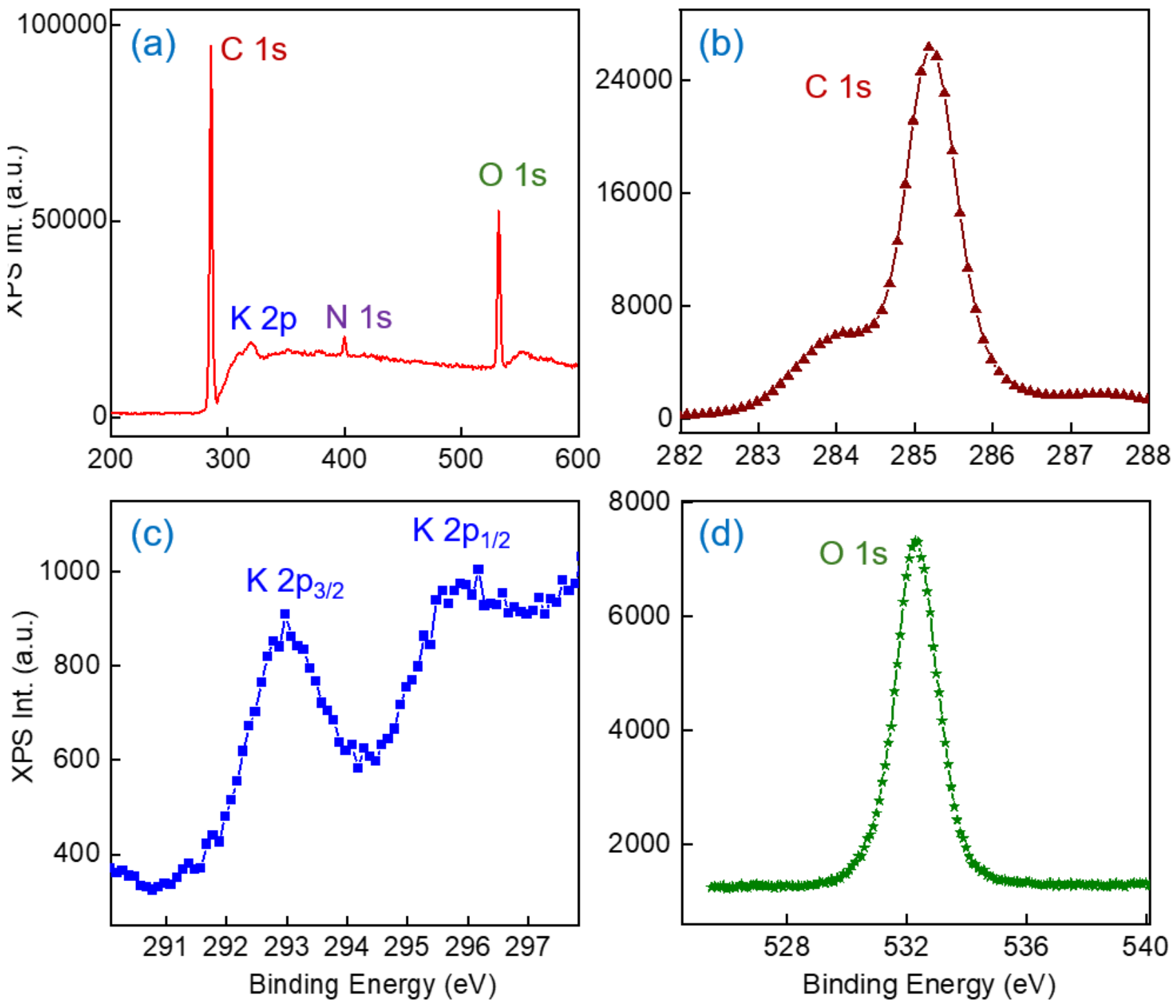

**Figure S11: (a)** Full XPS spectrum of the diamond surface, just after flowing the $KNO_3$ solution. High resolution XPS peaks with the background for C 1s (b), K 2p (c), O1s (d), and N 1s (e).